\documentclass[11pt,epsf,amssymb,qsymbols]{article}
\usepackage{tabularx}
\usepackage{array}
\usepackage{graphics}
\usepackage{graphicx}
\usepackage{psfrag}
\usepackage{epsfig}
\usepackage{amsmath}
\usepackage{amssymb}
\usepackage{ulem}
\usepackage{setspace}
\usepackage{rotating}
\usepackage{colortbl}
\usepackage{tabularx}
\usepackage{longtable}
\usepackage{multirow}
\usepackage[bookmarks=false,colorlinks]{hyperref}
\hypersetup{urlcolor=blue, citecolor=blue}
\makeatletter

\usepackage{verbatim}

\setlongtables

\newcommand{\bea}{\begin{eqnarray}}
\newcommand{\eea}{\end{eqnarray}}
\newcommand{\beq}{\begin{equation}}
\newcommand{\eeq}{\end{equation}}
\newcommand{\ec}{\end{center}}
\newcommand{\bc}{\begin{center}}
\newcommand{\gev}{{\rm GeV}}
\newcommand{\mev}{{\rm MeV}}

\newcommand{\pdir}{p\kern -5.2pt\raise 0.2ex\hbox {/}}

\newcommand{\vdir}{v\kern -5.75pt\raise 0.15ex\hbox {/}}
\newcommand{\kdir}{k\kern -5.75pt\raise 0.15ex\hbox {/}}
\newcommand{\epsdir}{\epsilon\kern -5.0pt\raise 0.15ex\hbox {/}}
\newcommand{\Ddir}{D\kern -7.75pt\raise 0.20ex\hbox {/}}
\newcommand{\Adir}{A\kern -7.75pt\raise 0.20ex\hbox {/}}
\newcommand{\ldir}{l\kern -5.0pt\raise 0.2ex\hbox{/}}
\newcommand{\varepsdir}{\varepsilon\kern -5.5pt\raise 0.15ex\hbox{/}}

\newcommand{\cm}{{\cal M}}
\def\at2{A_T^{(2)}(q^2)}
\newcommand{\atre}{A_T^{({\rm re})}(q^2)}
\newcommand{\atim}{A_T^{({\rm im})}(q^2)}

\def \eff{{\text{eff}}}

\def\I{\mathcal{I}}

\newcommand{\nn}{\nonumber}
\makeatother

\begin{document}
\begin{flushright}
\begin{tabular}{l}
{\small \tt LPT 12-79 }\\
{\small \tt LAL 12-274}
\end{tabular}
\end{flushright}
\begin{center}
\vskip 1.2cm\par
{\par\centering \textbf{\LARGE  
\Large \bf Impact of $B\to K^\ast_0 \ell^+\ell^-$ on the New Physics search  }}\\
\vskip .25cm\par
{\par\centering \textbf{\LARGE  
\Large  in $B\to K^\ast \ell^+\ell^-$ decay }}\\
\vskip 0.7cm\par
{\scalebox{.92}{\par\centering \large  
\sc Damir Be\v cirevi\'c$^a$ and Andrey Tayduganov$^{a,b}$ }}
{\par\centering \vskip 0.3 cm\par}
{\sl \small
$^a$~Laboratoire de Physique Th\'eorique (B\^at.~210)~\footnote{Laboratoire de Physique Th\'eorique est une unit\'e mixte de recherche du CNRS, UMR 8627.}\\
Universit\'e Paris Sud, Centre d'Orsay, F-91405 Orsay Cedex, France}\\
{\par\centering \vskip 0.3 cm\par}
{\sl \small
$^b$~Laboratoire de l'Acc\'el\'erateur Lin\'eaire, Centre d'Orsay, Universit\'e de Paris-Sud XI,\\
 B.P. 34,  B\^atiment 200, 91898 Orsay Cedex, France}\\
{\vskip 1.0cm\par}
\end{center}
\begin{abstract}
We discuss the uncertainty related to the amount of unwanted $B\to K_0^\ast (K\pi)\ell^+\ell^-$ events in the sample of $B\to K^\ast (K\pi)\ell^+\ell^-$ ones. Those events can increase the measured differential decay rate by up to $10\%$ in the low $q^2$ region, and can be a source of non-negligible uncertainty in the full angular distribution of the $B\to K^\ast (K\pi)\ell^+\ell^-$  decay. Although the transverse asymmetries should be unaffected by the presence of the $S$-wave $K\pi$ pairs, coming from the scalar $K_0^\ast$ meson, we show that in practice, their normalization might be sensitive to those events and could entail a sizable uncertainty in transverse asymmetries around $q^2=2~\gev^2$. For other $q^2$'s that error is under $10\%$.
\end{abstract}
\vskip 1.6cm
%\vskip 2.2 cm 
%\newpage
%\setcounter{page}{1}
%\setcounter{footnote}{0}
\setcounter{equation}{0}
\setcounter{footnote}{0}
%%%%%%%%%%%  Section 1
\section{Introduction}
The angular spectra of $B\to K^\ast (\to K\pi) \ell^+\ell^-$ decay contain a number of interesting observables which can be measured and compared with theoretical predictions made in the Standard Model (SM), and hopefully result in valuable hints about physics beyond the Standard Model (BSM).~\footnote{Recent research on the potential of angular distribution of the $B\to K^\ast \ell^+\ell^-$ decay in the search for signals of physics BSM can be found in ref.~\cite{theo} and references therein. The work in this direction was initiated by the authors of ref.~\cite{Melikhov}.}
To achieve that goal the main obstacle is our inability to reliably compute the hadronic quantities to desired accuracy. In such a situation one selects the observables that are the least sensitive to hadronic uncertainties that are at the same time potentially sensitive to the new physics (NP) signals.  It appears that three transverse asymmetries, that one can build up from the $q^2$-dependent coefficient functions in the analysis of the angular distribution of $B\to K^\ast (\to K\pi) \ell^+\ell^-$, satisfy these requirements at low $q^2$'s and are supposed to be well measured at LHCb as well as at Super-B and Super-KEKB/Belle~II (see also ref.~\cite{exp}). 

The asymmetry $\at2$  has been first introduced in ref.~\cite{Kruger:2005ep} and by now it is in the catalog of standard quantities that are expected to probe the effects of physics BSM at low energies. Its most significant feature is that 
\bea
\lim_{q^2\to 0}\at2 = {2{\rm Re} [C_7 C_7^{\prime \ast}] \over |C_7|^2 + |C_7^\prime |^2}\,,
\eea 
so that its non-zero value would suggest that $C_7^\prime \neq 0$, which could only be attributed to physics BSM. Away from $q^2=0$ this quantity is sensitive to other sources of NP and studying its shape at low $q^2$'s could be very beneficial for either seeing the effects BSM  or constraining the NP models. 

The asymmetry $\atim$ has been introduced in ref.~\cite{Egede:2008uy,Becirevic:2011bp} and it is particularly sensitive to the NP phases. In particular its intercept 
\bea
\lim_{q^2\to 0}\atim = {2\ {\rm Im} [C_7 C_7^{\prime \ast}] \over |C_7|^2 + |C_7^\prime |^2}\,,
\eea
and away from $q^2=0$ it is sensitive to the phases coming from operators other then the electromagnetic penguin operator. 

The asymmetry $\atre$ has been introduced  in ref.~\cite{Becirevic:2011bp} and is proportional to the usual forward-backward asymmetry $A_{FB}(q^2)$, except that it is not divided by the differential decay rate, but only by its transverse part for which the hadronic uncertainties are better controlled. 
It has the same zeros as $A_{FB}(q^2)$, and  a well established shape in the SM. In particular the position of its low $q^2$ extremum is well defined in terms of Wilson coefficients. More specifically, 
\bea\label{eq:rere}
&& {\rm for}~~ q_0^2  = - {2m_b\over R} {C_7\over C_9}, \qquad A_T^{({\rm re})}(q_0^2) = 0,  \ \cr
&&{\rm for}~~ q_{\rm extr}^2  = {2m_b\over R} {C_7\over C_{10}- C_9},  \qquad \biggl.\left( \partial \atre /\partial q^2\right)\biggr|_{q^2=q^2_{\rm extr}} \!\!\!\!=0\,,
\eea
so that from the position of its non-trivial zero and extremum at low $q^2$, one can already test the SM prediction~\footnote{$R$ is a ratio of form factors that will be defined below, which at low $q^2$'s behave like a constant. More specifically,  $R= {(A_1(q^2)/T_2(q^2))/( m_B - m_{K^\ast})} \simeq {(V(q^2)/T_1(q^2))/(m_B + m_{K^\ast})} \in (0.17, 0.23)\ \gev^{-1}$. See discussion in ref.~\cite{Becirevic:2011bp}.  }
\bea
{q_0^2\over q_{\rm extr}^2} = 1-  {C_{10}\over C_9}\,.
\eea 
The shape of $\atre$ is particularly sensitive to $C_7^{(\prime)}$. 

Most of the observed events so far are $B\to K^\ast \mu^+\mu^-$. As we are interested in exploring the low $q^2$ region it is important to include the lepton mass effects, as to properly combine $B\to K^\ast \mu^+\mu^-$ with $B\to K^\ast e^+ e^-$ events. 
A problem that is often ignored in the literature is the contamination of the angular distribution of $B\to K^\ast (\to K\pi) \ell^+\ell^-$ by the events coming from 
$B\to K_0^\ast (\to K\pi) \ell^+\ell^-$, where $K_0^\ast$ stands for a broad scalar meson resonance. This effect was recently studied in the experimental analysis of e.g. $D\to K^\ast \mu \nu$ decay~\cite{scalarr1, scalarr2, scalarr3}, and was shown to be important.  For the above mentioned asymmetries this is not a problem because the product  of the $K_0^\ast \to K\pi$ decay is in its $S$-wave and cannot make any impact on the $B\to K^\ast (\to K\pi) \ell^+\ell^-$ transverse amplitudes. However, in the extraction of transverse asymmetries from the full angular distribution the unwanted $(K\pi)_S$ originating from $B\to K^\ast_0\ell^+\ell^-$ are troublesome and result in an error that is $q^2$-dependent and can be uncomfortably large, as we show in the following. One should also mention that for very low $q^2$ a special care should be devoted to the presence of the light resonances $\rho^0$, $\omega$ and $\phi$, the effect of which should be subtracted away. 

In Sec.~\ref{sec:2} we derive expressions for the full distribution of the combined $B\to K^\ast (\to K\pi)\ell^+\ell^-$ and $B\to K^\ast_0 (\to K\pi) \ell^+\ell^-$ decays, and in Sec.~\ref{sec:3} we discuss the phenomenological consequences.  

\section{$B\to K^\ast(\to K\pi) \ell^+\ell^-$ and $B\to K_0^\ast(\to K\pi) \ell^+\ell^-$\label{sec:2}}

\subsection{Operator basis and the hadronic matrix elements}
The effective Hamiltonian used to describe the $b\to s\ell^+\ell^-$ decay reads~\cite{Heff},
 \begin{equation} \label{eq:Heff}
  {\cal H}_{\eff} = - \frac{4\,G_F}{\sqrt{2}} V_{tb}V_{ts}^\ast
     \left[  \sum_{i=1}^{6} C_i (\mu)
\mathcal O_i(\mu) + \sum_{i=7,8,9,10} \biggl(C_i (\mu) \mathcal O_i + C'_i (\mu) \mathcal
O'_i\biggr)\right] \,,
\end{equation}
where the twice Cabibbo suppressed contributions ($\propto  V_{ub}V_{us}^\ast $) have been neglected and the operator basis in which the Wilson coefficients have been computed is~\cite{Bobeth:1999mk,Altmannshofer:2008dz}:
\begin{align}
{\mathcal{O}}_{7} &= \frac{e}{g^2} m_b
(\bar{s} \sigma_{\mu \nu} P_R b) F^{\mu \nu} ,&
{\mathcal{O}}_{7}^\prime &= \frac{e}{g^2} m_b
(\bar{s} \sigma_{\mu \nu} P_L b) F^{\mu \nu} ,\nn \\
{\mathcal{O}}_{9} &= \frac{e^2}{g^2} 
(\bar{s} \gamma_{\mu} P_L b)(\bar{\ell} \gamma^\mu \ell) ,&
{\mathcal{O}}_{9}^\prime &= \frac{e^2}{g^2} 
(\bar{s} \gamma_{\mu} P_R b)(\bar{\ell} \gamma^\mu \ell) ,\nn \\
{\mathcal{O}}_{10} &=\frac{e^2}{g^2}
(\bar{s}  \gamma_{\mu} P_L b)(  \bar{\ell} \gamma^\mu \gamma_5 \ell) ,&
{\mathcal{O}}_{10}^\prime &=\frac{e^2}{g^2}
(\bar{s}  \gamma_{\mu} P_R b)(  \bar{\ell} \gamma^\mu \gamma_5 \ell) ,
 \end{align}
with $P_{L,R}=(1\mp \gamma_5)/2$. The explicit expressions for $\mathcal O_{1-6}$ and $\mathcal O_{8}^{(\prime)}$ can be found in ref.~\cite{Bobeth:1999mk}, together with the Wilson coefficients $C_i (\mu)$ which carry the SM information on physics at short distances. The Wilson coefficients multiplying the same hadronic matrix element are combined into effective coefficients~\cite{Wilsoneff} and in what follows whenever we write $C_{7,9,10}$, the effective coefficients $C_{7,9,10}^\eff$ should be understood. Note also that in the SM the Wilson coefficients $C_{7-10}^\prime =0$.

The hadronic matrix elements, obtained by sandwiching the above operators between $B$ and $K^\ast$ and/or between $B$ and $K^\ast_0$, describe the long distance physics and are parameterized in terms of the form factors. There are seven form factors parameterizing the $B\to K^\ast $ transition  matrix elements, namely, 
\bea\label{eq:ffVA}
\langle K^\ast (k,\varepsilon) | \bar s\gamma_\mu(1-\gamma_5) b | B(p)\rangle  &=&  \varepsilon_{\mu\nu\rho\sigma}\varepsilon^{*\nu} p^\rho k^\sigma\,
\frac{2V(q^2)}{m_B+m_{K^\ast}} - i \varepsilon^\ast_\mu (m_B+m_{K^\ast}) A_1(q^2)\nonumber \\
&& \hspace*{-32mm}+ i (p+k)_\mu (\varepsilon^\ast \cdot q)\, \frac{A_2(q^2)}{m_B+m_{K^\ast}}   +  i q_\mu (\varepsilon^\ast \cdot q) \,\frac{2m_{K^\ast}}{q^2}\,
\left[A_3(q^2)-A_0(q^2)\right], 
\eea
where the partial conservation of the axial current provides the following relation,
\bea
A_3(q^2) & = & \frac{m_B+m_{K^\ast}}{2m_{K^\ast}}\, A_1(q^2) -
\frac{m_B-m_{K^\ast}}{2m_{K^\ast}}\, A_2(q^2)\,,
\eea
that cancels the divergence at $q^2=0$, via the condition $A_0(0)=A_3(0)$.
Other three form factors parameterize the matrix element of the electromagnetic penguin operator, 
\bea\label{eq:ffT}
\langle K^\ast(k,\varepsilon) | \bar s \sigma_{\mu\nu} q^\nu (1+\gamma_5) b |
B(p)\rangle &=& 2 i\varepsilon_{\mu\nu\rho\sigma} \varepsilon^{*\nu}
p^\rho k^\sigma \,  T_1(q^2)\nonumber\\
& & + \left[ \varepsilon^\ast_\mu
  (m_B^2-m_{K^\ast}^2) - (\varepsilon^\ast \cdot q) \,(p+k)_\mu \right] T_2(q^2) \nonumber\\
& &+  
(\varepsilon^\ast \cdot q) \left[ q_\mu - \frac{q^2}{m_B^2-m_{K^\ast}^2}\, (p+k)_\mu
\right]T_3(q^2),
\eea
with $T_1(0) = T_2(0)$, ensuring that only one  form factor describes the physical $B\to K^\ast \gamma$ decay.  Concerning the decay to the scalar meson, the analogous matrix elements involve three new form factors, namely,
\bea\label{eq:ffS}
&&\langle   K_0^\ast(k)\vert \bar s\gamma_\mu \gamma_5 b \vert B(p)\rangle = \left[ (p+k)_\mu - {m_B^2-m_{K_0^\ast}^2\over q^2} q_\mu  \right] f_+(q^2) +  {m_B^2-m_{K_0^\ast}^2\over q^2} q_\mu f_0(q^2)\,,\nn\\
 &&  \langle   K_0^\ast(k)\vert \bar s \sigma_{\mu\nu} \gamma_5 q^\nu b \vert B(p)\rangle = i\left[  (m_B^2-m_{K_0^\ast}^2) q_\mu - q^2 (p+k)_\mu   \right] { f_T(q^2)\over m_B+m_{K_0^\ast}}\,.
\eea

\subsection{$B\to K^\ast \ell^+\ell^-$ and $B\to K_0^\ast \ell^+\ell^-$ amplitudes}

With the above definitions in hands we can now write the $B\to K^\ast \ell^+\ell^-$ decay amplitudes as:
\begin{align}\label{eq:wu0}
	\cm_{\perp}^{L,R} =& - N_1\sqrt{ 2 \lambda_{K^\ast} N_{K^\ast}} \left[ (C_9^{(+)}\mp C_{10}^{(+)}) {V(q^2)  \over m_B+m_{K^\ast} }  +   C_7^{(+)}  {2 m_b \over q^2 }  T_1(q^2)
\right]\,,\nn\\
	&\cr
	\cm_{\parallel}^{L,R} =&  N_1\sqrt{ 2 N_{K^\ast}} ( m_B^2-m_{K^\ast}^2 ) \left[  (C_9^{(-)}\mp C_{10}^{(-)}) { A_1(q^2) \over  m_B-m_{K^\ast} }  +   C_7^{(-)}  {2 m_b \over q^2 } T_2(q^2)
\right]\,,\nn\\
	&\cr
	\cm_{0}^{L,R} =&  {N_1\sqrt{N_{K^\ast}} \over 2m_{K^\ast}\sqrt{q^2} }  \left\{  (C_9^{(-)}\mp C_{10}^{(-)}) \left[ ( m_B^2-m_{K^\ast}^2 -q^2)(m_B+m_{K^\ast}) A_1(q^2) -  {\lambda_{K^\ast} \over m_B+m_{K^\ast}} A_2(q^2)\right]\right. \nn\\
	&\left. +2 m_b C_7^{(-)} \left[ (m_B^2+3 m_{K^\ast}^2-q^2)T_2(q^2) -  {\lambda_{K^\ast} \over m_B^2-m_{K^\ast}^2} T_3(q^2) \right]\right\}\,,\nn\\
	&\cr
	\cm_{t}^{L,R} =&  {N_1\sqrt{\lambda_{K^\ast} N_{K^\ast}} \over \sqrt{q^2} }  (C_9^{(-)}\mp C_{10}^{(-)})A_0(q^2) \,,
\end{align}
where, for shortness, we write $C_{7,9,10}^{(\pm)} = C_{7,9,10} \pm  C_{7,9,10}^{ \prime}$, and~\footnote{The above amplitudes are related to $A_{\perp,\parallel,0,t}^{L,R}(q^2)$, defined in  
ref.~\cite{Altmannshofer:2008dz}, as: $\cm_{\perp,\parallel,0,t}^{L,R} (q^2)= -i\sqrt{{3/8}} \  A_{\perp,\parallel,0,t}^{L,R}(q^2)$.
} 
\bea
N_1 &=& i {G_F\over\sqrt{2}} {\alpha_{\rm em}\over 4\pi} V_{tb}V^\ast_{ts}\,,\quad N_{K_{\rm res}} = {q^2\beta_\ell \sqrt{\lambda_{K_{\rm res}}} \over 256\pi^3 m_B^3}\,,\quad\beta_\ell=\sqrt{1-{4m_\ell^2\over q^2}}\,,  \nn\\
 \quad   \lambda_{K_{\rm res}} &\equiv & \lambda(m_B^2, m_{K_{\rm res}}^2, q^2)= [q^2 - (m_B+m_{K_{\rm res}})^2][q^2 - (m_B-m_{K_{\rm res}})^2]\,.
\eea
The above amplitudes agree with ref.~\cite{Lu:2011jm}. As for those describing $B\to K^\ast_0 \ell^+\ell^-$, they are given by:
\begin{align}\label{eq:wu1}
\cm_0^{L,R\,\prime}(q^2) =& i {N_1 \sqrt{N_{K_0^\ast} \lambda_{K_0^\ast}} \over \sqrt{q^2}} \left[ (C_9^{(-)}\mp C_{10}^{(-)}) f _+(q^2) +  C_7^{(-)} 2 m_b { f_T(q^2) \over m_B+m_{K_0^\ast}} \right]\,,\nn\\
\cm_t^{L,R\,\prime}(q^2) =& i {N_1 \sqrt{N_{K_0^\ast}} \over \sqrt{q^2}} (C_9^{(-)}\mp C_{10}^{(-)}) (m_B^2-m_{K_0^\ast}^2) f_0(q^2) \,,
\end{align}
where an extra ``prime" is used to distinguish them from the $B\to K^\ast$ amplitudes which also agree with ref.~\cite{Lu:2011jm}. The superscript $L,R$ in the above expressions refers to the lepton pair chirality. 

\begin{figure}[t!]
\psfrag{aa}{\color{blue}\Huge $q^2$} 
\psfrag{Standard}{\color{blue}\huge \hspace*{-30mm}$ $} 
\begin{center}
{\resizebox{8cm}{!}{\includegraphics{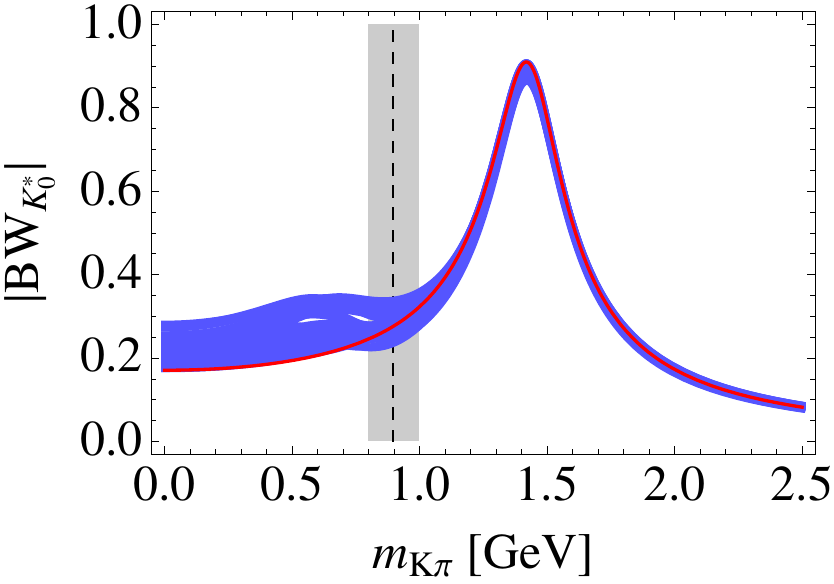}}}~{\resizebox{8cm}{!}{\includegraphics{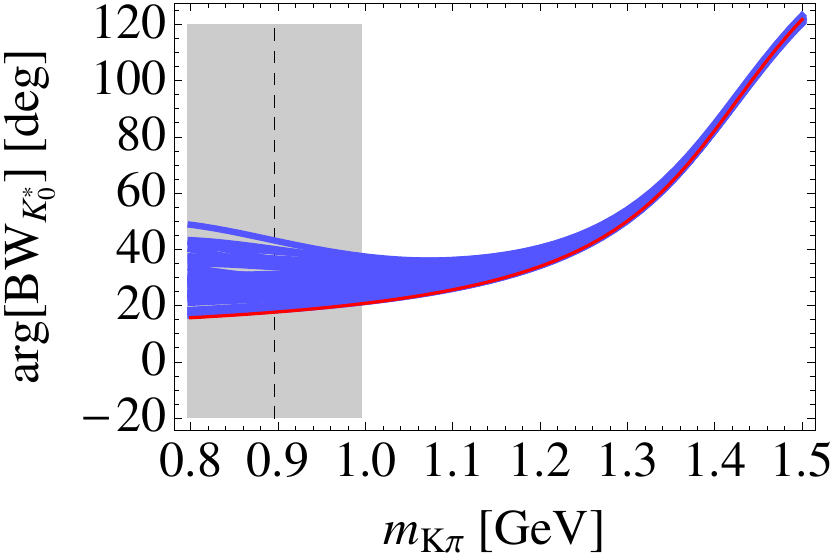}}} 
\caption{\label{fig:1}\footnotesize{\sl 
Moduli and the $S$-wave phase of the function $BW_{K_{(0)}^\ast}(m_{K\pi}^2)$ defined in eq.~(\ref{eq:bww}): the thick curve corresponds to eq.~(\ref{eq:bwk0star}) with $g_\kappa=0$, while the shaded area is obtained by accounting for the $\kappa$-state by varying $| g_\kappa |\in [0,0.2]$ and $\arg(g_\kappa) \in [\pi/2, \pi]$. The vertical stripe corresponds to $m_{K^\ast}\pm 100\ \mev$, which is the only part that is of interest for the subject of this paper.
}} 
\end{center}
\end{figure}

Finally, we should take into account the width of resonances, which we do by multiplying each amplitude by its corresponding Breit-Wigner function, namely
\bea
&& \cm_{\perp,\parallel,0,t}^{L,R}(q^2) \longrightarrow  \cm_{\perp,\parallel,0,t}^{L,R}(q^2) \ BW_{K^\ast}(m_{K\pi}^2)\,,\nn\\
&&  \hfill\nn\\ 
&& \cm_{0,t}^{L,R\ \prime}(q^2) \longrightarrow  \cm_{0,t}^{L,R\ \prime}(q^2) \  BW_{K_0^\ast}(m_{K\pi}^2)\,,
 \eea
where the functions $BW_{K_{(0)}^\ast}(m_{K\pi}^2)$ given by
\bea\label{eq:bww}
&&BW_{K^\ast}(m_{K\pi}^2) ={  \sqrt{m_{K^\ast}\Gamma_{K^\ast}/\pi} \over m_{K^\ast}^2 - m_{K\pi}^2 - i m_{K^\ast}\Gamma_{K^\ast}}\,,\\
&&\label{eq:bwk0star}BW_{K_0^\ast}(m_{K\pi}^2) ={\cal N} \left[ -{ g_\kappa \over (m_{\kappa} - i \Gamma_{\kappa}/2)^2 -  m_{K\pi}^2 } + { 1 \over (m_{K_0^\ast} - i \Gamma_{K_0^\ast}/2)^2 - m_{K\pi}^2 } \right]  \,,
\eea
and the factor ${\cal N}$ is obtained from the normalization to unity, 
\bea\label{eq:00}
\int_{-\infty}^{\infty} dm_{K\pi}^2\  |BW_{K_0^\ast}(m_{K\pi}^2)|^2 =1\,.
\eea
The second term in eq.~(\ref{eq:bwk0star}) is the contribution of the well measured $K_0^\ast=K_0^\ast(1430)$, the mass and width of which are $m_{K_0^\ast}=1425(50)$~MeV and $\Gamma_{K_0^\ast}=270(80)$~MeV, respectively. The first term in eq.~(\ref{eq:bwk0star}) accounts for the $K_0^\ast(800)\equiv \kappa$ state, identified as a pole 
on the second Riemann sheet in the  $K\pi\to K\pi$ scattering amplitude, with the mass and width being $m_\kappa =658(13)$~MeV and $\Gamma_\kappa=557(24)$~MeV, respectively~\cite{DescotesGenon:2006uk}. To mimic its presence in the tail of the Breit-Wigner function for $K^\ast_0$, we vary the moduli of the coupling $0 \lesssim |g_\kappa| \lesssim 0.2$, where the upper value should be conservative enough as it results in a prominent second bump in $BW_{K_0^\ast}(m_{K\pi}^2)$ (see fig.~\ref{fig:1}). If that bump was as large it would have been directly observed in experiments long before ref.~\cite{DescotesGenon:2006uk}. 
Concerning the phase of the parameter $g_\kappa$ we checked that its variation within $\arg(g_\kappa)\in [\pi/2,\pi]$, gives the overall $S$-wave phase of $BW_{K_0^\ast}(m_{K\pi}^2)$ that is compatible with experimental results of refs.~\cite{Aston:1987ir,scalarr3} in the region of $m_{K\pi} \in [ m_{K^\ast}-\delta, m_{K^\ast}+\delta ]$, where $\delta \approx 100$~MeV. The function~(\ref{eq:bwk0star}) also reproduces well the findings of refs.~\cite{ DescotesGenon:2006uk} and~\cite{scaFF}. The resulting $|BW_{K_0^\ast}(m_{K\pi}^2)|$ and $\arg(BW_{K_0^\ast}(m_{K\pi}^2))$ are plotted in fig.~\ref{fig:1}.  

We attempted varying the phase $\arg(g_\kappa)$ outside the region indicated above and found that: (a) for $\arg(g_\kappa) \in [0,\pi/2[$ the phase $\arg(BW_{K_0^\ast}(m_{K\pi}^2))$ remains lower than the one measured in refs.~\cite{Aston:1987ir,scalarr3}, (b) for  $\arg(g_\kappa) \in [\pi,2 \pi[$, the bump in $|BW_{K_0^\ast}(m_{K\pi}^2)|$ would become a dip, which would mean less events in the sample around the $\kappa$-states, which would be in conflict with experiments. 

Therefore, in the following we use the functions $BW_{K^\ast,K_0^\ast}(m_{K\pi}^2)$ written above with $g_\kappa = |g_\kappa| \exp[i\arg(g_\kappa)]$, and parameters  $|g_\kappa| \in [0,0.2]$, $\arg(g_\kappa)\in [\pi/2,\pi]$ tuned to reproduce the known experimental data for $(K\pi)_S$ around the $m_{K\pi}=m_{K^\ast}$, the region which is relevant to the subject of the present paper. 

\subsection{Full distribution of $B\to K^\ast(K\pi)\ell^+\ell^-$ including $B\to K_0^\ast(K\pi)\ell^+\ell^-$}
Besides the usual variables, $q^2$, $\theta_\ell$, $\theta_K$ and $\phi$, we now also need to consider the distribution of the mass of the $K\pi$ system. We have
\bea\label{eq:full}
&&\hspace*{-10mm}{d^5\Gamma\over dq^2 dm_{K\pi}^2 d\cos\theta_\ell d\cos\theta_K d\phi}= J_1^c(q^2, m_{K\pi}^2,\theta_K) + 2 J_1^s(q^2, m_{K\pi}^2,\theta_K) \nn\\
&&\hfill\cr
&&+ [J_2^c(q^2, m_{K\pi}^2,\theta_K)+ 2J_2^s(q^2, m_{K\pi}^2,\theta_K)]\cos2\theta_\ell+ 2 J_3(q^2, m_{K\pi}^2,\theta_K)\sin^2\theta_\ell \cos2\phi  \nn\\
&&\hfill\cr
&&+ 2 \sqrt{2} J_4(q^2, m_{K\pi}^2,\theta_K)\sin(2\theta_\ell)\cos\phi + 2\sqrt{2} J_5(q^2, m_{K\pi}^2,\theta_K)\sin\theta_\ell\cos\phi  \nn\\
&&\hfill\cr
&&+ 2J_6(q^2, m_{K\pi}^2,\theta_K)\cos\theta_\ell + 2\sqrt{2}J_7(q^2, m_{K\pi}^2,\theta_K)\sin\theta_\ell\sin\phi \nn\\
&&\hfill\cr
&&  + 2\sqrt{2} J_8(q^2, m_{K\pi}^2,\theta_K)\sin2\theta_\ell\sin\phi + 2J_9(q^2, m_{K\pi}^2,\theta_K)\sin^2\theta_\ell\sin2\phi ,
\eea
where the explicit forms of the functions $J_{1-9}^{(c,s)}(q^2,m_{K\pi}^2,\theta_K)$ look as follows:
\begin{align}
J_{1,2}^s(q^2, m_{K\pi}^2,\theta_K) =& {3 \over 8\pi}\I_{1,2}^s(q^2)\vert BW_{K^\ast}(m_{K\pi}^2)\vert^2 \sin^2\theta_K \,,\nn
\end{align}
\begin{align}
J_{3,6,9}(q^2, m_{K\pi}^2,\theta_K) =& {3 \over 8\pi} \I_{3,6,9}(q^2)\vert BW_{K^\ast}(m_{K\pi}^2)\vert^2 \sin^2\theta_K\,,\nn
\end{align}
\begin{align}
J_1^c(q^2, m_{K\pi}^2,\theta_K) =& {1\over 4\pi}\biggl\{\I_1^{c\,\prime}(q^2)\vert BW_{K_0^\ast}(m_{K\pi}^2)\vert^2 + 3 \I_1^c(q^2)\vert BW_{K^\ast}(m_{K\pi}^2)\vert^2\cos^2\theta_K \biggr. \nn\\
	& \biggl. + 2\sqrt3{\rm Re}\left[ \I_1^{c\,\prime\prime}(q^2) BW_{K_0^\ast}(m_{K\pi}^2) BW_{K^\ast}^\dagger(m_{K\pi}^2) \right] \cos\theta_K \biggr\} \,,\nn\\
	&\cr
J_2^c(q^2, m_{K\pi}^2,\theta_K) =& {1 \over 4\pi}\biggl\{\I_2^{c\,\prime}(q^2)\vert BW_{K_0^\ast}(m_{K\pi}^2)\vert^2 + 3\I_2^c(q^2)\vert BW_{K^\ast}(m_{K\pi}^2)\vert^2 \cos^2\theta_K \biggr.\nn\\
	&\biggl. +2 \sqrt3{\rm Re}\left[ \I_2^{c\,\prime\prime}(q^2) BW_{K_0^\ast}(m_{K\pi}^2) BW_{K^\ast}^\dagger(m_{K\pi}^2) \right]\cos\theta_K \biggr\} \,,\nn\end{align}
\begin{align}
J_4(q^2, m_{K\pi}^2,\theta_K) =& \sqrt{\frac{3}{2}} \frac{1}{4\pi} \biggl\{ {\rm Re}\left[\I_4^{\,\prime\prime}(q^2)BW_{K_0^\ast}(m_{K\pi}^2) BW_{K^\ast}^\dagger(m_{K\pi}^2) \right]\sin\theta_K \biggr. \nn\\
	& \biggl. +{\sqrt3 \over 2} \I_4(q^2)\vert BW_{K^\ast}(m_{K\pi}^2)\vert^2\sin2\theta_K \biggr\} \,,\nn	\end{align}
\begin{align}
J_5(q^2, m_{K\pi}^2,\theta_K) =& \sqrt{\frac{3}{2}} \frac{1}{4\pi} \biggl\{ {\rm Re}\left[\I_5^{\,\prime\prime}(q^2)BW_{K_0^\ast}(m_{K\pi}^2) BW_{K^\ast}^\dagger(m_{K\pi}^2) \right]\sin\theta_K \biggr. \nn\\
	& \biggl. +{\sqrt3 \over 2} \I_5(q^2)\vert BW_{K^\ast}(m_{K\pi}^2)\vert^2\sin2\theta_K \biggr\} \,,\nn
	\end{align}
\begin{align}
J_7(q^2, m_{K\pi}^2,\theta_K) =& \sqrt{\frac{3}{2}} \frac{1}{4\pi} \biggl\{ {\rm Im}\left[\I_7^{\,\prime\prime}(q^2)BW_{K_0^\ast}(m_{K\pi}^2) BW_{K^\ast}^\dagger(m_{K\pi}^2) \right]\sin\theta_K \biggr. \nn\\
	& \biggl. +{\sqrt3 \over 2} \I_7(q^2)\vert BW_{K^\ast}(m_{K\pi}^2)\vert^2\sin2\theta_K \biggr\} \,,\nn\\
	&\cr
J_8(q^2, m_{K\pi}^2,\theta_K) =& \sqrt{\frac{3}{2}} \frac{1}{4\pi} \biggl\{ {\rm Im}\left[\I_8^{\,\prime\prime}(q^2)BW_{K_0^\ast}(m_{K\pi}^2) BW_{K^\ast}^\dagger(m_{K\pi}^2) \right]\sin\theta_K \biggr. \nn\\
	& \biggl. +{\sqrt3 \over 2} \I_8(q^2)\vert BW_{K^\ast}(m_{K\pi}^2)\vert^2\sin2\theta_K \biggr\} \,.
\end{align}
The functions $\I_i^{(s,c)}(q^2)$, expressed in terms of the amplitudes $\cm_{\perp,\parallel,0,t}^{L,R} (q^2)$ from eq.~(\ref{eq:wu0}), are listed in the appendix of the present paper, together with the functions  $\I_i^{(c)\,\prime}(q^2)$ 
 that involve the scalar meson contributions only~(\ref{eq:wu1}), and the functions  $\I_i^{(c)\,\prime\prime}(q^2)$ that correspond to the interference terms.~\footnote{The functions $\I_i^{(s,c)}(q^2)$ are related to  the familiar $I_i^{(s,c)}(q^2)$ of ref.~\cite{Altmannshofer:2008dz} as: $\I_i^{(s,c)}(q^2)=({3/8})\ I_i^{(s,c)}(q^2)$.}

The transversity amplitudes obviously remain unchanged when the $K\pi$ from $K^\ast_0$ are included in the decay's angular distribution. Only the scalar and the ``$t$" amplitudes get changed. Since we focus on the transverse asymmetries only, we could have ignored the whole issue. However, since the transverse asymmetries are (implicitly) normalized to the differential $B\to K^\ast \ell^+\ell^-$ decay rate in a given $q^2$-bin, it is important to estimate the size of the contribution of the $K\pi$-pairs originating from $K^\ast_0$, relative to the dominant $K^\ast \to K\pi$, inside the window around the $K^\ast$ mass.

\subsection{Separate distributions in $\phi$, $\theta_\ell$, $\theta_K$}

We now integrate the full distribution~(\ref{eq:full}) to get the dependence of $d^2\Gamma/dq^2dm_{K\pi}^2$ on the angles $\phi$, $\theta_\ell$, $\theta_K$ separately. 
Above and in the following $\Gamma$ stands for the decay width comprising both $B\to K^\ast  (\to K\pi)\ell^+\ell^-$ and $B\to  K_0^\ast  (\to K\pi)\ell^+\ell^-$ modes. 
We get: 
\begin{align} 
{d^3\Gamma \over dq^2dm_{K\pi}^2 d \cos\theta_K}& =  \frac{1}{3} \biggl\{ \bigl[ 3 \I_1^{c\,\prime}(q^2) - \I_2^{c\,\prime}(q^2) \bigr]\ \vert BW_{K_0^\ast}(m_{K\pi}^2)\vert^2 \biggr. \nn\\
	&  + 3 \bigl[ 3 \I_1^s(q^2) - \I_2^s(q^2) \bigr]\ \vert BW_{K^\ast}(m_{K\pi}^2)\vert^2\biggr.\nn\\
	&   + 2\sqrt{3} {\rm Re} \bigl[ \left( 3\I_1^{c\,\prime\prime}(q^2) - \I_2^{c\,\prime\prime}(q^2) \right) BW_{K_0^\ast}(m_{K\pi}^2) BW_{K^\ast}^\dag(m_{K\pi}^2) \bigr] \cos\theta_K\nn\\
	& \biggl. +3 \bigl[ 3\I_1^c(q^2)- \I_2^c(q^2) - 3\I_1^s(q^2)+ \I_2^s(q^2) \bigr] \vert BW_{K^\ast}(m_{K\pi}^2)\vert^2 \cos^2\theta_K\biggr\}\,,\label{eq:K} 
\end{align}
\begin{align} 
{d^3\Gamma  \over dq^2dm_{K\pi}^2 d\cos \theta_\ell} & = \bigl[ \I_1^{c\,\prime}(q^2) - \I_2^{c\,\prime}(q^2) \bigr] \vert BW_{K_0^\ast}(m_{K\pi}^2) \vert^2 \nn\\
	 & + \bigl[ \I_1^{c}(q^2) - \I_2^{c}(q^2) + 2 \I_1^{s}(q^2) - 2 \I_2^{s}(q^2) \bigr]  \vert BW_{K^\ast}(m_{K\pi}^2)\vert^2 \nn\\
	 & + 2 \I_6(q^2) \vert BW_{K^\ast}(m_{K\pi}^2) \vert^2 \cos\theta_\ell \nn\\
	 & + 2\bigl[ \I_2^{c\,\prime}(q^2) \vert BW_{K_0^\ast}(m_{K\pi}^2)\vert^2 + \left( \I_2^c(q^2) + 2 \I_2^s(q^2)\right) \vert BW_{K^\ast}(m_{K\pi}^2) \vert^2 \bigr] \cos^2\theta_\ell \,,\label{eq:ell}
\end{align}
\begin{align} 
{d^3\Gamma  \over dq^2dm_{K\pi}^2 d\phi} & =  \frac{1}{3\pi } \biggl\{ \bigl[ 3  \I_1^{c\,\prime}(q^2) - \I_2^{c\,\prime}(q^2) \bigr] \vert BW_{K_0^\ast}(m_{K\pi}^2) \vert^2\biggr.\nn\\
	& + \bigl[ 3 \I_1^{c}(q^2) -\I_2^c(q^2) + 6 \I_1^{s}(q^2)  - 2 \I_2^s(q^2) \bigr] \vert BW_{K^\ast}(m_{K\pi}^2) \vert^2
 \nn\\
	&  + {3\sqrt{3}\pi^2\over 8} {\rm Re}\bigl[ \I_5^{\,\prime\prime}(q^2)BW_{K_0^\ast}(m_{K\pi}^2) BW_{K^\ast}^\dag (m_{K\pi}^2)\bigr] \cos\phi \nn\\
	&  + {3\sqrt{3}\pi^2\over 8} {\rm Im} \bigl[ \I_7^{\,\prime\prime}(q^2) BW_{K_0^\ast}(m_{K\pi}^2) BW_{K^\ast}^\dag (m_{K\pi}^2) \bigr] \sin\phi \nn\\
	&\biggl. + 4\bigl[ \I_3(q^2) \cos 2\phi + \I_9(q^2) \sin 2\phi \bigr] \vert BW_{K^\ast}(m_{K\pi}^2) \vert^2 \biggr\}\,.\label{eq:phi}
\end{align}

\begin{figure}[t!]
\psfrag{aa}{\color{blue}\Huge $q^2$} 
\psfrag{Standard}{\color{blue}\huge \hspace*{-30mm}$ $} 
\begin{center}
{\resizebox{10.7cm}{!}{\includegraphics{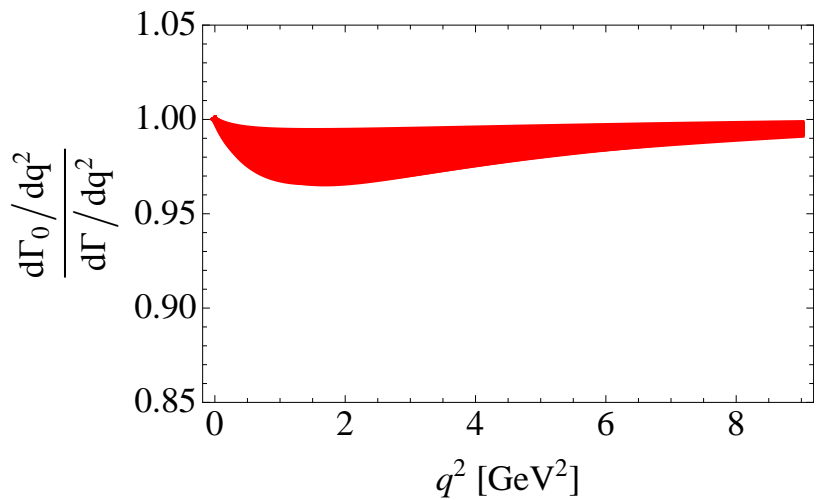}}} 
\caption{\label{fig:2}\footnotesize{\sl 
Ratio between $d\Gamma_0/dq^2$, the $B\to K^\ast (\to K\pi)\mu^+\mu^-$ differential decay rate, and $d\Gamma/dq^2$ that includes also the $B\to K^\ast_0 (\to K\pi)\mu^+\mu^-$ events integrated over the interval $m_{K\pi}\in [m_{K^\ast} - 100~\mev, m_{K^\ast} + 100~\mev ]$. We limit the discussion to low $q^2 < m_{J/\psi}^2$.
}} 
\end{center}
\end{figure}

\noindent To get the differential decay width we integrate over all three angles and obtain
\bea\label{eq:fullwidth}
{d^2\Gamma\over dq^2dm_{K\pi}^2} &=& \frac{2}{3} \biggl\{  \bigl[ 3 \I_1^{c\,\prime}(q^2) - \I_2^{c\,\prime}(q^2)\bigr]\ \vert BW_{K_0^\ast}(m_{K\pi}^2)\vert^2 \biggr.  \nn\\
&&\qquad \biggl. +\bigl[ 3 \I_1^{c}(q^2) -\I_2^c(q^2) + 6 \I_1^{s}(q^2) - 2 \I_2^s(q^2) \bigr] \vert BW_{K^\ast}(m_{K\pi}^2) \vert^2  \biggr\} \nn\\
&=&  {d^2\Gamma_{S}\over dq^2dm_{K\pi}^2} + {d^2\Gamma_{0}\over dq^2dm_{K\pi}^2}\,,
\eea
where we separated the part coming from the scalar resonance ($\Gamma_S$) from the usual expression for $B\to K^\ast \ell^+\ell^-$ decay ($\Gamma_0$). We stress again that when integrating over $m_{K\pi}^2 \in [ (m_{K^\ast}-\delta)^2 , (m_{K^\ast}+\delta )^2  ]$, besides the desired $K\pi$ corresponding to $K^\ast$, one also gets the $K\pi$ pairs coming from $K_0^\ast$. In particular, in fig.~\ref{fig:2} we show the ratio of $d\Gamma_0/dq^2$ and $d\Gamma/dq^2$, after integrating over $m_{K\pi}^2$ using $\delta=100$~MeV, as in experiments.~\footnote{The plots presented in this work are obtained by treating the uncertainties to all the form factors and to $g_\kappa$ parameter as uniform distributions, while those related to masses and widths of the states are treated as Gaussian.} We see that the inclusion of $K\pi$ coming from $K_0^\ast$ amounts to at most  $5\%$ excess with respect to the desired $d\Gamma_{0}/dq^2$. One should keep in mind that for this estimate: (i) we used the form factors from ref.~\cite{ball-zwicky} obtained by using QCD sum rules on the light cone, an approximate method the uncertainties of which are hard to assess; (ii) we also had to use the form factors $A_{2,0}(q^2)$, $T_3(q^2)$, that are prone to large uncertainties~\cite{lattV}; (iii) for the $B\to K^\ast_0$ transition form factors we used the results of ref.~\cite{aliev}, obtained from the standard QCD sum rule analysis of the three-point correlation functions; (iv) we varied  the parameter in eq.~(\ref{eq:bwk0star}), that enters also eq.~(\ref{eq:fullwidth}), as $|g_\kappa| \in [0,0.2]$, $\arg(g_\kappa)\in [\pi/2,\pi]$.  Note also that the normalization condition in eq.~(\ref{eq:00}) is an assumption that could be checked by a careful comparison of the $D\to K^\ast$ and $D\to K_0^\ast$ semileptonic form factors computed in the narrow resonance approximation on the lattice with the experimental results presented in ref.~\cite{scalarr3}.

\section{Phenomenology\label{sec:3}}
\subsection{Error on transverse asymmetries}
Expressed in terms of the functions $\I_i^{(s,c)}(q^2)$, the three transverse asymmetries mentioned in introduction,  are: 
\begin{align}\label{eq:at}
\at2 = {4 \I_3(q^2)\over 3 \I_1^s(q^2)- \I_2^s (q^2)}&\,,\quad \atim = {4 \I_9(q^2)\over 3 \I_1^s(q^2)- \I_2^s (q^2)}\,,\nn\\
 \atre = &{\beta_\ell \I_6^s(q^2)\over 3 \I_1^s(q^2)- \I_2^s (q^2)}\,.\quad 
\end{align}
Throughout this paper we keep the lepton mass different from zero because the lowest bins (as close to $q^2\approx 0$ as possible) are the least ambiguous as far as the identification of the potential NP signal is concerned, and in those bins the effect of the lepton mass is significant. Moreover, in order to consistently combine the samples of $B\to K^\ast e^+e^-$ and $B\to K^\ast \mu^+\mu^-$ decays at low $q^2$'s the lepton mass  effect should be taken into account . 

In the massless lepton case the denominators in eq.~(\ref{eq:at}) reduce to $3 \I_1^s(q^2)- \I_2^s (q^2) = 8 \I_2^s (q^2)$ and one retrieves the usual expressions quoted in e.g. refs.~\cite{Altmannshofer:2008dz}. A common denominator to all three asymmetries is chosen for convenience; it consists of  the transverse amplitudes only (those that are unaffected by the presence of $K\pi$ pairs coming from $K_0^\ast$). In principle, and if the sample of $B\to K^\ast \ell^+\ell^-$ is so large that all the coefficient functions $\I_i^{s,c}(q^2)$ can be extracted from the full angular distribution~(\ref{eq:full}), one can get the denominator unaffected by the presence of the $K\pi$ pairs coming from $K_0^\ast$ decay. If, instead, one proceeds by considering the distribution of the sample of $B\to K^\ast \ell^+\ell^-$ events in $\phi$, $\theta_\ell$, and $\theta_K$ separately, then the denominator $3 \I_1^s(q^2)- \I_2^s (q^2)$ cannot be extracted without picking up the events coming from $B\to K^\ast_0 (\to K\pi)\ell^+\ell^-$ decay. To be more specific, we rewrite eq.~(\ref{eq:K}) as
\begin{figure}[t!]
\psfrag{aa}{\color{blue}\Huge $q^2$} 
\psfrag{Standard}{\color{blue}\huge \hspace*{-30mm}$ $} 
\begin{center}
{\resizebox{10.2cm}{!}{\includegraphics{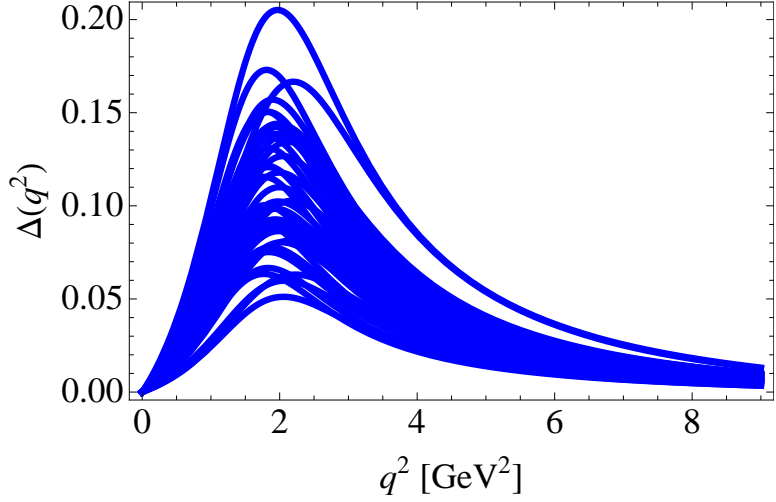}}} 
\caption{\label{fig:3}\footnotesize{\sl 
$\Delta(q^2)$ is a contribution to $a_{\theta_K}(q^2)$ in the distribution (\ref{eq:bk}) of $B\to K^\ast (\to K\pi)\ell^+\ell^-$ coming from the $B\to K^\ast_0 (\to K\pi)\ell^+\ell^-$ decay.
}} 
\end{center}
\end{figure}
\bea\label{eq:bk}
{d^2\Gamma\over dq^2 d\cos\theta_K} = a_{\theta_K} (q^2)+ b_{\theta_K} (q^2)\cos\theta_K + c_{\theta_K} (q^2)\cos^2\theta_K\,, 
\eea
where the coefficient functions are easily identified from eq.~(\ref{eq:K}). For our purpose, the important one is 
\begin{align}
a_{\theta_K}(q^2) &= \frac{1}{3}  \bigl[ 3 \I_1^{c\,\prime}(q^2) - \I_2^{c\,\prime}(q^2) \bigr]\int dm_{K\pi}^2 \vert BW_{K_0^\ast}(m_{K\pi}^2)\vert^2 \nn\\
& \qquad +   \bigl[ 3 \I_1^s(q^2) - \I_2^s(q^2) \bigr]\int dm_{K\pi}^2 \vert BW_{K^\ast}(m_{K\pi}^2)\vert^2 \nn\\
				   &= \bigl[ 3 \I_1^s(q^2) - \I_2^s(q^2) \bigr] \bigl[1+\Delta(q^2) \bigr] \int dm_{K\pi}^2 \vert BW_{K^\ast}(m_{K\pi}^2)\vert^2\ ,
\end{align}
where, as before, the integration over $m_{K\pi}^2$ comprises $(m_{K^\ast}-\delta)^2 \leq m_{K\pi}^2\leq (m_{K^\ast}+\delta)^2$. Without the $B\to K^\ast_0 (\to K\pi)\ell^+\ell^-$ events the functions $\I_{1,2}^{c\,\prime}(q^2) =0$, and therefore $a_{\theta_K} (q^2)$ would be the way to fix the denominators in eq.~(\ref{eq:at}). 
Since this is not the case, and the functions  $\I_{1,2}^{c\,\prime}(q^2) \neq 0$,  the function $\Delta(q^2)\neq 0$. By using the same form factors as discussed above we obtain $\Delta(q^2)$ shown in fig.~\ref{fig:3}, where the band of values covers the uncertainties coming from all the form factors, as well as from the variation of the phenomenological constant $g_\kappa$, as discussed after eq.~(\ref{eq:00}). Note that in this estimate the errors on $B\to K^\ast$ form factors are small as only the transverse amplitudes are considered in this case, which do not involve $A_{2,0}(q^2)$ not $T_3(q^2)$ form factors~\cite{Becirevic:2011bp}. From  the plot in fig.~\ref{fig:3} we see that around $q^2 \approx 2~\gev^2$ as many as $20\ \%$ events, recognized as $a_{\theta_K} (q^2)$ of $B\to K^\ast \ell^+\ell^-$, might be coming from $B\to K_0^\ast (\to K\pi)\ell^+\ell^-$ decay. There is nothing extraordinary about $q^2\approx 2\ \gev^2$; it is simply a point at which the denominator of $\Delta(q^2)$ has a minimum, while its numerator depends only mildly on $q^2$.  We also checked that the contamination of the $B_s\to\phi\,\ell^+\ell^-$ decay by the $B_s\to f_0(980)\ell^+\ell^-$ events affects $a_{\theta_K} (q^2)$ by an excess that is similar in shape to the one shown in fig.~\ref{fig:3} except that its maximum at about $q^2\approx 2.5\ \gev^2$ is no larger than $0.12$. For that estimate we used $\delta=10$~MeV (as used by LHCb~\cite{LHCbphi}), and the $B_s\to f_0$ form factors computed in ref.~\cite{defazz}.

As for the numerators in $\at2$ and $\atim$ they are {\sl not} plagued by $B\to K_0^\ast (\to K\pi)\ell^+\ell^-$, and they are extracted from eq.~(\ref{eq:phi}) written as,
\bea\label{eq:bphi}
{d^2\Gamma\over dq^2 d\phi} = a_\phi (q^2) + b^c_\phi (q^2) \cos\phi + b^s_\phi (q^2) \sin\phi  +  c^c_\phi (q^2) \cos 2\phi + c^s_\phi (q^2) \sin2\phi\,, 
\eea  
after identifying $4\I_3(q^2)= 3 \pi c^c_\phi (q^2)$, and $4\I_9(q^2)= 3 \pi c^s_\phi (q^2)$. Note also that $2 \pi a_\phi(q^2) = d\Gamma/dq^2$. Concerning the numerator of $\atre$, it is extracted in the usual way from eq.~(\ref{eq:ell}), 
\bea
{d^2\Gamma\over dq^2 d\cos\theta_\ell} = a_{\theta_\ell} (q^2)+ b_{\theta_\ell} (q^2)\cos\theta_\ell + c_{\theta_\ell} (q^2)\cos^2\theta_\ell\,, 
\eea 
with $\I_6(q^2) = b_{\theta_\ell}/2$. 

In concluding this subsection, we repeat that the $B\to K_0^\ast (\to K\pi)\ell^+\ell^-$ events are unavoidable when studying the transverse asymmetries from the angular distribution of  $B\to K^\ast (\to K\pi)\ell^+\ell^-$ decay, due to the practical impossibility of extracting the denominators in eq.~(\ref{eq:at}) without getting any contributions coming from the functions  $\I_{i}^{c\,\prime}(q^2)$. Any other choice of normalizing the transverse amplitudes in eq.~(\ref{eq:at}) would be as much (if not more) affected by the $B\to K_0^\ast (\to K\pi)\ell^+\ell^-$ events. The only way to circumvent the problem discussed above is to consider the full distribution~(\ref{eq:full}) and extract the coefficient functions $\I_i^{s,c}(q^2)$ from the simultaneous fit of the angular dependence in $\theta_K$, $\theta_\ell$ and $\phi$. In doing so one would have to work in the massless lepton limit, which means for $q^2 > 1\ \gev^2$.

\subsection{The interference terms}

\begin{figure}[t!]
\hspace*{-11mm}{\resizebox{9cm}{!}{\includegraphics{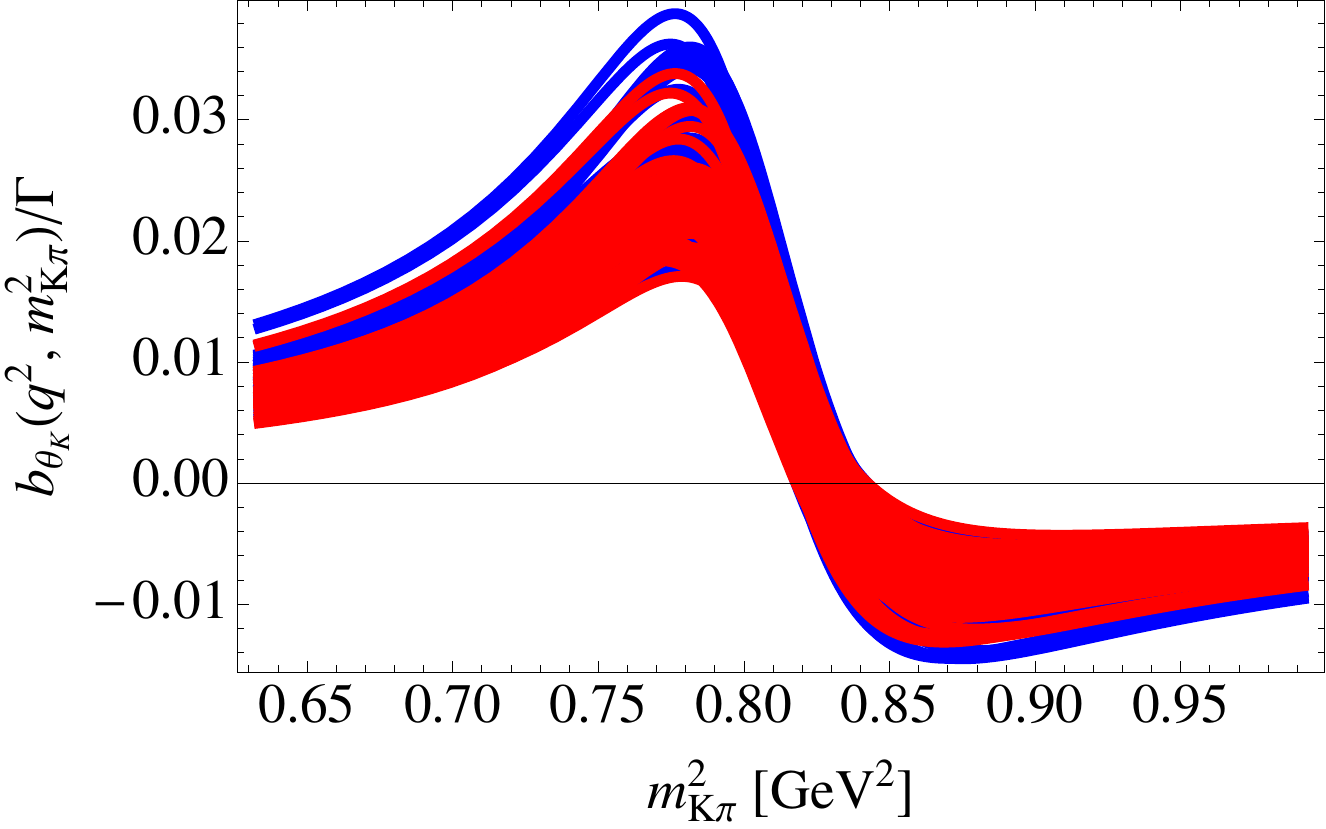}}}~{\resizebox{9cm}{!}{\includegraphics{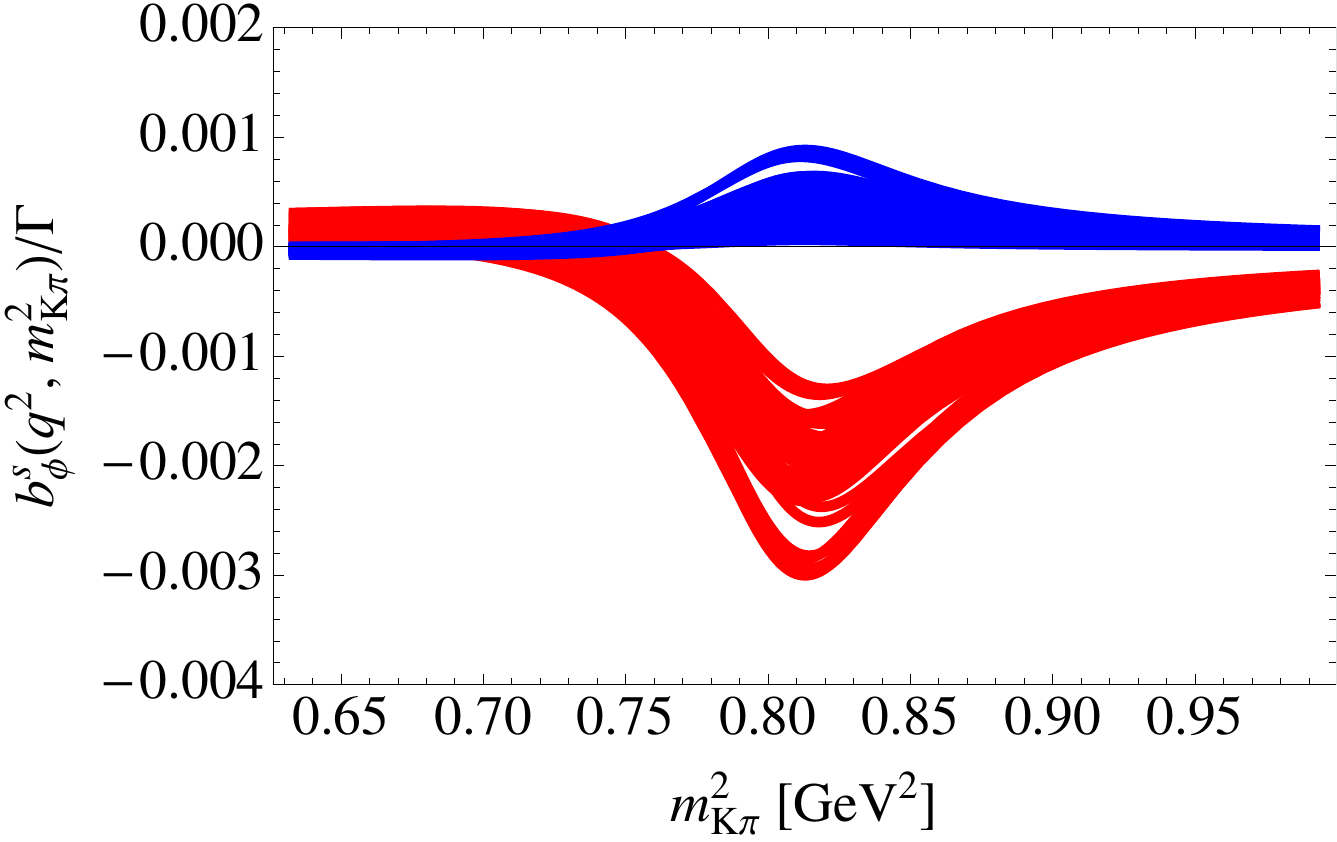}}}  
\caption{\label{fig:4}\footnotesize{\sl 
Dependence of the interference terms on the $K\pi$ mass around $m_{K^\ast}$ for $q^2=1 \ \gev^2$ (blue), and $q^2=5 \ \gev^2$  (red). Left plot corresponds to $b_{\theta_K}(q^2,m_{K\pi}^2)$ which, when integrated over  $m_{K\pi}^2$, leads to $b_{\theta_K}(q^2)$ in eq.~(\ref{eq:bk}). Similarly, in the right plot we show  $b_{\phi}^{s}(q^2,m_{K\pi}^2)$  that, after the integration in $m_{K\pi}^2$ between $(m_{K^\ast} -\delta)^2$ and $(m_{K^\ast} +\delta)^2$, gives the function $b_{\phi}^{s}(q^2)$ used in eq.~(\ref{eq:bphi}). For convenience, both functions are normalized to the total decay width.}} 
\end{figure}

It is interesting to comment on the terms arising from interference between the $B\to K^\ast (\to K\pi)\ell^+\ell^-$  and the $B\to K_0^\ast (\to K\pi)\ell^+\ell^-$  amplitudes, that we denoted as $b_{\theta_K}(q^2)$ in eq.~(\ref{eq:bk})  [explicitly given in eq.~(\ref{eq:K})], and by $b_{\phi}^{s,c}(q^2)$ in eq.~(\ref{eq:bphi})  [and explicitly in eq.~(\ref{eq:phi})]. Before integrating in $m_{K\pi}^2$ the functions  $b_{\theta_K}(q^2,m_{K\pi}^2)$ and  $b_{\phi}^c(q^2,m_{K\pi}^2)$ are nearly antisymmetric and symmetric with respect to $m_{K^\ast}^2$ (c.f.  fig.~\ref{fig:4}). This is  a consequence of the fact that they are proportional to the real part of the Breit-Wigner function $BW_{K^\ast}(m_{K\pi}^2)$ which changes the sign at $m_{K^\ast}^2$. The contribution of $BW_{K^\ast_0}(m_{K\pi}^2)$ is too small to make a significant impact when integrating over  $m_{K\pi}^2 \in [(m_{K^\ast}-\delta)^2, (m_{K^\ast}+\delta)^2]$ with $\delta \approx 100\ \mev$, and therefore  $b_{\theta_K}(q^2)$ and  $b_{\phi}^c(q^2)$ are not exactly zero but very close to it. The non-vanishing interference term is $b_{\phi}^{s}(q^2)$, i.e. the one that involves the imaginary part of the Breit-Wigner function, as shown in fig.~\ref{fig:4}. The (anti-)symmetry between the regions $m_{K\pi}<m_{K^\ast}$ and $m_{K\pi}>m_{K^\ast}$ is mostly due to the phase $\arg(g_\kappa) \neq 0$.

\begin{figure}[t!]
\begin{center}{\resizebox{10.4cm}{!}{\includegraphics{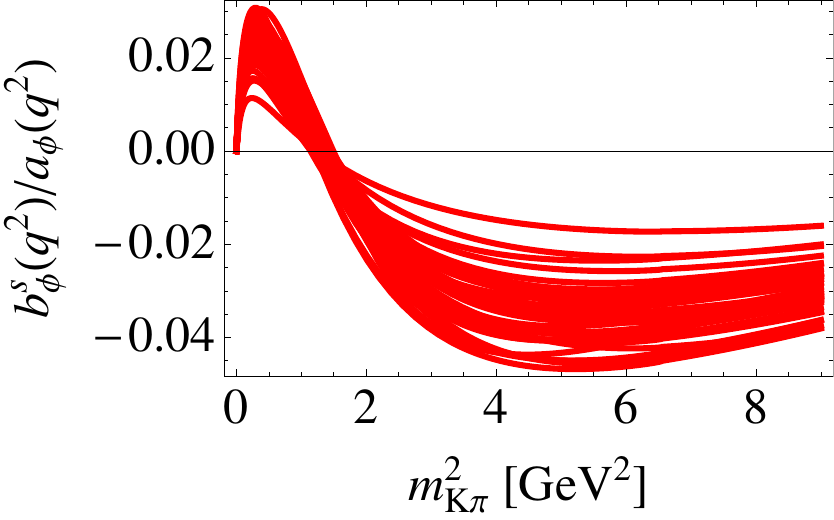}}}  
\end{center}
\caption{\label{fig:5}\footnotesize{\sl Dependence of the interference term $b_{\phi}^{s}(q^2)$ (divided by the positive definite function $a_{\phi}(q^2)$) on $q^2$. It crosses zero at $q_{\rm int}^2= 1.3(3)~\gev^2$, that coincides with the low $q^2$ extremum of $\atre$, indicated in eq.~(\ref{eq:rere}). 
}} 
\end{figure}

An interesting feature of the function $b_{\phi}^{s}(q^2)$ is that it has zero at the same point at which the asymmetry $\atre$ reaches its extremum. To see that it suffices to write 
\bea
b_{\phi}^{s}(q^2) &=& {\sqrt{3}\pi\over 8} \int dm_{K\pi}^2\  {\rm Im} \left[ \I_7^{\prime\prime}(q^2) BW_{K_0^\ast}(m_{K\pi}^2)BW_{K^\ast}^\ast(m_{K\pi}^2)\right]\nn\\
&\propto & {\rm Re} \left[ \cm_0^{L\,\prime} \cm_\parallel^{L \ast} - \cm_0^{R\,\prime} \cm_\parallel^{R \ast} \right] \approx  {\rm Re} \left[ \cm_0^{L\,\prime} \cm_\parallel^{L \ast} \right] \propto 
 {\rm Re}\  \cm_\parallel^{L \ast}  \nn\\
&\propto&    {\rm Re}\left[ (C_9^{(-)} - C_{10}^{(-)}) { A_1/T_2 \over  m_B-m_{K^\ast} }  +   C_7^{(-)}  {2 m_b \over q^2 }\right] \,,
\eea
which is zero for 
\bea\label{eq:qint}
q_{\rm int}^2 ={2 m_b \over R}  {{\rm Re}\ C_7^{(-)} \over {\rm Re} [  C_{10}^{(-)} - C_9^{(-)}]} \,,
\eea 
as illustrated in fig.~\ref{fig:5}. Therefore a study of the interference term $b_{\phi}^{s}(q^2)$ can be informative. In practice, the $b_{\phi}^{s}(q^2)$ signal can be made more significant if a larger value of $\delta$ is chosen. 

\section{Summary}
In this paper we discussed the impact of $B\to K_0^\ast (\to K\pi)\ell^+\ell^-$ events on the angular distribution of the $B\to K^\ast (\to K\pi)\ell^+\ell^-$ decay. A study of the latter is expected to lead to important clues about the potential signals of physics BSM. In particular the transverse asymmetries are supposed to be unaffected by the $S$-wave $K\pi$ pairs. We show that in practice, however, the extraction of these asymmetries at low $q^2$'s is plagued by the $B\to K^\ast_0 (\to K\pi)\ell^+\ell^-$ events and that the corresponding error is under $10\%$ for $q^2\lesssim 1~\gev^2$  and for  $4\ \gev^2 \lesssim q^2 < m_{J/\psi}^2$, while it might be as large as $20\%$ around $q^2\approx 2\ \gev^2$.  Similar situation occurs in the $B_s\to \phi (\to K^+K^-)\ell^+\ell^-$ decay except that the effect of the $B_s\to f_0 (\to K^+K^-)\ell^+\ell^-$ decay is smaller. It remains under $12\%$ around $q^2\approx 2.5\ \gev^2$, and under $5\%$ elsewhere. To arrive to such a conclusion we considered the dependencies on the angles $\theta_K$, $\theta_\ell$ and $\phi$ separately, i.e. in the way in which the three transverse asymmetries could be extracted if the $B\to K_0^\ast \ell^+\ell^-$ events were absent. 
 At large $q^2\gtrsim 14 \ \gev^2$, instead, the effect of  $B\to K_0^\ast  \ell^+\ell^-$ and $B_s\to f_0(980) \ell^+\ell^-$ on $B\to K^\ast  \ell^+\ell^-$ and $B_s\to \phi \ell^+\ell^-$ respectively, is completely negligible.  
   
This uncertainty, together with the one related to the charm loop~\cite{Khodjamirian:2010vf}, and controllable uncertainties on the ratios of the $B\to K^\ast$ form factors ($R$), suggests that the overall error on the transverse asymmetries~(\ref{eq:at}) is under about $30\%$, and therefore at that level of accuracy their measurement remains a good tool for detecting the NP signal.  This uncertainty can be circumvented only if the coefficient functions, appearing in the full distribution~(\ref{eq:full}), are extracted from the simultaneous fit to the angular dependence of the decay events in three angles $\theta_\ell$, $\theta_K$ and $\phi$, which is a problem that could hopefully be solved experimentally once the number of events becomes sufficiently large for such a fit to be reliable. 

\vspace*{31mm}

\section*{Acknowledgments}
We thank Jacques Lefran\c{c}ois for discussions on the experimental aspects of $B\to K^\ast \ell^+\ell^-$, S\'ebastien Descotes-Genon for the discussion on the $\kappa$ state, as well as Wei Wang and Joaquim Matias for correspondance. Research by A.T. has been partly supported by {\sl Agence Nationale de la Recherche}, contract LFV-CPV-LHC ANR-NT09-508531.

\newpage 
\appendix

\section{Explicit expressions for $\I_{1-9}^{(c,s,\prime,\prime\prime)}(q^2)$ }
In this appendix we collect the formulas that relate the coefficient functions $\I_{1-9}^{(c,s)}(q^2)$ appearing in the distributions~(\ref{eq:K},\ref{eq:ell},\ref{eq:phi}) with the $B\to K^\ast$ amplitudes $ \cm_{\perp,\parallel,0,t}^{L,R}(q^2)$ given in eq.~(\ref{eq:wu0}), and with the $B\to K^\ast_0\ell^+\ell^-$ amplitudes  $\cm_{0,t}^{L,R\ \prime}(q^2)$ given in eq.~(\ref{eq:wu1}). We have:
\bea\label{distr-3}
&&\I_1^s(q^2)   =   \frac{2 + \beta^2_\ell}{4} \left[ \lvert \cm_{\perp}^L
	\rvert ^2 + \lvert \cm_{\parallel}^L  \rvert ^2 + \left( L \rightarrow R
	\right)  \right] + \frac{4m_\ell^2}{q^2} \text{Re} \left( \cm_{\parallel}^L 
	\cm_{\parallel}^{R \ast}  + \cm_{\perp}^L  \cm_{\perp}^{R \ast } \right)\,,\nn
	\\
&&\I_1^{c\,(\prime)}(q^2)  = \lvert \cm_{0}^{L\,(\prime)}\rvert^2 + \lvert \cm_{0}^{R\,(\prime)} \rvert^2 +
	\frac{4 m_{\ell}^2}{q^2} \left[ \lvert \cm_t^{\,(\prime)} \rvert^2 +2 \text{Re} \left(
	\cm_{0}^{L\,(\prime)} \cm_{0}^{R\,(\prime)\ast} \right) \right] \,,\nn \\
&&\I_2^s(q^2)  = \frac{\beta_\ell^2}{4} \left[ \lvert \cm_{\perp}^L \rvert ^2
	+ \lvert \cm_{\parallel}^L \rvert ^2 + \left( L \rightarrow R \right) 
	\right] \,,\nn\\
&&\I_2^{c\,(\prime)}(q^2)  = -\beta_\ell^2 \left( \lvert \cm_{0}^{L\,(\prime)}\rvert^2 + \lvert
	\cm_{0}^{R\,(\prime)} \rvert^2 \right)\,, \nn\\
&&\I_3(q^2)  = \frac{\beta_\ell^2}{2} \left[ \lvert \cm_{\perp}^L\rvert ^2 -
	\lvert \cm_{\parallel}^L \rvert ^2 +  \left( L \rightarrow R \right) 
	\right] \,,\nn\\
&&\I_4(q^2)  = \frac{\beta_{\ell}^2}{\sqrt{2}} \left[ \text{Re} \left(
	\cm_{0}^L \cm_{\parallel}^{L *} \right) + \left( L \rightarrow R \right) 
	\right] \,,\nn \\
&&\I_5(q^2)  = \sqrt2 \beta_\ell \left[ \text{Re} \left( \cm_{0}^L
	\cm_{\perp}^{L \ast} - \left( L \rightarrow R \right)  \right)  \right]\,, \nn\\
&&\I_6(q^2)  = 2 \beta_\ell \left[ \text{Re}\left( \cm_{\parallel}^L
	\cm_{\perp}^{L \ast} \right) - \left( L \rightarrow R \right) \right] \,,
	\nn\\
&&	\I_7(q^2) = \sqrt{2} \beta_\ell \left[ \text{Im} \left(\cm_{0}^L
	\cm_{\parallel}^{L  \ast} - \left( L \rightarrow R \right)  \right)  \right] \,, \nn\\
&&\I_7(q^2) = \sqrt{2} \beta_\ell \left[ \text{Im} \left(\cm_{0}^L
	\cm_{\parallel}^{L  \ast} - \left( L \rightarrow R \right)  \right)  \right] \,,\nn\\
&&\I_8(q^2)  = \frac{\beta_\ell^2}{\sqrt2} \left[ \text{Im} \left(
	\cm_{0}^L \cm_{\perp}^{L *} \right) +  \left( L \rightarrow R \right) \right]
	\,,\nn\\
&& \I_9(q^2)     = \beta_\ell^2\,\left[ \text{Im} \left( \cm_{\perp}^L
	\cm_{\parallel}^{L \ast} \right) + \left( L \rightarrow R \right) \right]\,.
\eea
Note that the above functions $\I_i^{(s,c)}(q^2)$ are related to $I_i^{(s,c)}(q^2)$ of ref.~\cite{Altmannshofer:2008dz} as $\I_i^{(s,c)}(q^2)=({3/8})\ I_i^{(s,c)}(q^2)$. 
Concerning the interference terms in eqs.~(\ref{eq:K},\ref{eq:phi}) they read: 
\bea
&&\I_1^{c\,\prime\prime}(q^2) = \cm_0^{L\,\prime} \cm_0^{L \ast} + \cm_0^{R\,\prime} \cm_0^{R \ast}
	+ {4m_\ell^2\over q^2} \left[ \cm_0^{L\,\prime} \cm_0^{R \ast}+ \cm_0^L \cm_0^{R\,\prime \ast}+ \cm_t^{\,\prime} \cm_t^{ \ast}
\right] \,,\nn\\
&&\I_2^{c\,\prime\prime}(q^2)= -\beta_\ell^2 \bigl[ \cm_0^{L\,\prime} \cm_0^{L \ast} + \cm_0^{R\,\prime} \cm_0^{R \ast}\bigr]  \,,\nn\\
&&\I_4^{\,\prime\prime}(q^2)= \frac{\beta_\ell^2}{\sqrt2} \bigl[ \cm_0^{L\,\prime} \cm_\parallel^{L \ast} + \cm_0^{R\,\prime} \cm_\parallel^{R \ast} \bigr] \,, \nn\\
&&\I_5^{\,\prime\prime}(q^2)= \sqrt{2}\beta_\ell \bigl[ \cm_0^{L\,\prime} \cm_\perp^{L \ast} - \cm_0^{R\,\prime} \cm_\perp^{R \ast} \bigr] \,,\nn\\
&&\I_7^{\,\prime\prime}(q^2)= \sqrt{2}\beta_\ell \bigl[ \cm_0^{L\,\prime} \cm_\parallel^{L \ast} - \cm_0^{R\,\prime} \cm_\parallel^{R \ast} \bigr] \,,\nn\\
&&\I_8^{\,\prime\prime}(q^2)= \frac{\beta_\ell^2}{\sqrt2} \bigl[ \cm_0^{L\,\prime} \cm_\perp^{L \ast} + \cm_0^{R\,\prime} \cm_\perp^{R \ast} \bigr] \,.
\eea
For shortness, in expression for $\I_1^{c\,\prime\prime}(q^2)$ we used  $ \cm_t^{(\prime)} =  \cm_t^{R(\prime)} -  \cm_t^{L(\prime)}$.

\end{document}